# Unbinding of Retinoic Acid from its Receptor Studied by Steered Molecular Dynamics

Dorina Kosztin, Sergei Izrailev, and Klaus Schulten
Departments of Chemistry and Physics, Beckman Institute, University of Illinois at Urbana-Champaign, Urbana, Illinois 61801 USA

ABSTRACT   Retinoic acid receptor (RAR) is a ligand-dependent transcription factor that regulates the expression of genes involved in cell growth, differentiation, and development. Binding of the retinoic acid hormone to RAR is accompanied by conformational changes in the protein which induce transactivation or transrepression of the target genes. In this paper we present a study of the hormone binding/unbinding process in order to clarify the role of some of the amino acid contacts and identify possible pathways of the all-*trans* retinoic acid binding/unbinding to/from human retinoic acid receptor (hRAR)-γ. Three possible pathways were explored using steered molecular dynamics simulations. Unbinding was induced on a time scale of 1 ns by applying external forces to the hormone. The simulations suggest that the hormone may employ one pathway for binding and an alternative "back door" pathway for unbinding.

## INTRODUCTION

Retinoic acid (RA) is an important regulator of cellular proliferation and differentiation in higher eukaryotes. The action of RA is mediated by two types of nuclear hormone receptors: the retinoic acid receptor (RAR) and the retinoid X receptor (RXR). They act as transcriptional enhancers that bind to specific sequences of DNA and activate transcription after the ligand bound to the receptor induces conformational changes. Both RAR and RXR are members of the nuclear hormone receptor family and share a modular structure composed of several domains, each performing a specific function (Krust et al., 1986; Kumar et al., 1987; Freedman and Luisi, 1993). The C-terminal domain, also called the ligand binding domain (LBD), is responsible for recognizing and binding the hormone as well as for controlling multiple receptor functions such as transactivation and transrepression (Wagner et al., 1995; Bourguet et al., 1995; Renaud et al., 1995; Judelson and Privalsky, 1996).

Crystal structures of the ligand binding domain of the apo human retinoid-X receptor RXR-α (Bourguet et al., 1995), the human retinoic acid receptor hRAR-γ bound to all-*trans* retinoic acid (Renaud et al., 1995) and to 9-*cis* retinoic acid (Klaholz et al., 1998), the rat α₁ thyroid hormone receptor (TR) bound to a thyroid hormone agonist (Wagner et al., 1995), the estrogen receptor (ER) bound to agonist and antagonist (Brzozowski et al., 1997; Tanenbaum et al., 1998), and progesterone complexed to its receptor (PR) (Williams and Sigler, 1998) have been resolved recently and have provided insight into the mechanism of hormone binding.

A comparison of the available crystal structures of the LBD of different members of the nuclear hormone receptor family indicates that a conformational change in the receptor accompanies the transition between the liganded and the unliganded forms. This conformational change, also anticipated by a variety of indirect experiments involving alteration in receptor hydrophobicity, heat shock protein binding, thermal stability, and protease sensitivity (Driscoll et al., 1996), enables the hormone-receptor complex to bind to specific sequences of DNA as well as to other transcriptional co-activator or co-repressor proteins (Brent et al., 1989; Damm et al., 1989; Andersson et al., 1992). Elucidation of the mechanism by which hormone binding induces conformational changes is critical for understanding the functional role of the hormone.

Unfortunately, no crystal structures of the same receptor with bound and unbound ligand are presently available, and comparison of the LBD structures of different receptors may not reveal the conformational changes induced by ligand binding for a single receptor. It is still uncertain how large the conformational changes induced by hormone binding are, and what they entail. The two C-terminal helices, H11 and H12, were experimentally proven to be involved in hormone binding (Lee et al., 1995; Martinez et al., 1997; Vombaur et al., 1998). However, it is still not clear what role helix H12 plays during the hormone binding. Renaud et al. (1995) proposed that when t-RA binds to the receptor, helix H12 changes its conformation from extended into solvent for the structure of hRXR-α (Fig. 1 *b*), to a conformation closing the entrance to the binding pocket for hRAR-γ (Fig. 1 *a*). It has also been proposed (Renaud et al., 1995) that the hormone attracted by residue Lys-264 "drags" helix H12 in the folded conformation on its way toward the binding pocket, and that the salt bridge between residues Glu-414 and Lys-264 is responsible for locking helix H12 in this position. However, the crystal structure of the estrogen receptor bound to agonist also shows helix H12 in the extended conformation, but this positioning is an artefact of the crystal packing (Tanenbaum et al., 1998). Since the continuous α-helix H12 in the RXR structure has

## FIGURES

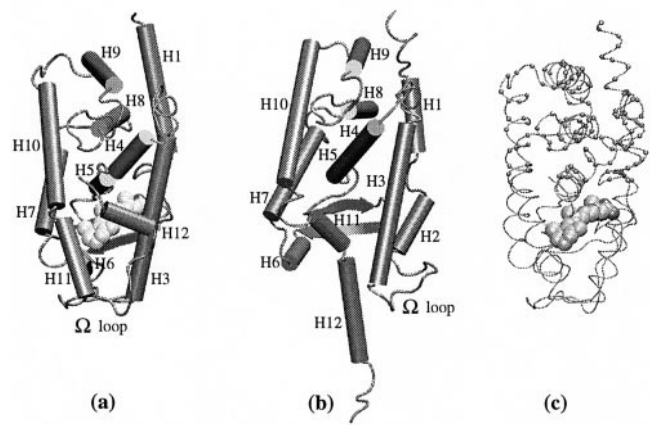

FIGURE 1 (*a*) Crystal structure of the human retinoic acid receptor (hRAR)-γ bound to all-*trans* retinoic acid (Renaud et al., 1995); (*b*) crystal structure of the human retinoid-X receptor RXR-α (Bourguet et al., 1995) showing an extended conformation of helices H11 and H12, as well as the Ω-loop. The LBD, in both structures, forms an antiparallel α-helical sandwich with a three layer structure: helices H4, H5, H6, H8, and H9 enclosed by helix H1 and H3 on one side and helices H7, H10, and H11 on the other side [for the topology diagram of the secondary structure see Renaud et al. (1995)]. (*c*) Simulated hRAR-γ–t-RA system; all α-carbon atoms that were restrained during the forced unbinding are represented as small spheres.

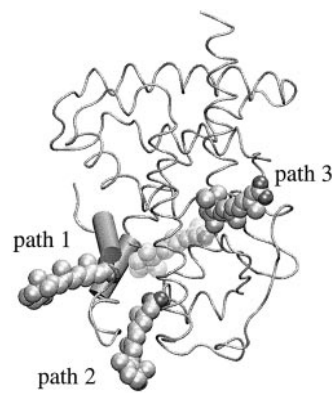

FIGURE 2 The unbinding of the t-RA hormone (initial position shown as transparent vdW spheres) was enforced along three pathways. Path 1: the hormone is pulled out between helices H11 and H12 (represented as cylinders); path 2: the hormone is pulled out beneath helices H11 and H12; path 3: the hormone is pulled out through the only discernible opening in the protein surface.

a rather low helical propensity for the corresponding sequence, the positioning of this helix may also be a result of crystal packing. The structures mentioned reveal that the receptors undergo a reorganization of the tertiary structure upon hormone binding (Renaud et al., 1995; Wagner et al., 1995; Brzozowski et al., 1997); however, this reorganization varies for each protein and the differences among the crystal structures may or may not be related to hormone binding. Hence, in addition to the structural analysis, investigation of ligand-receptor binding/unbinding dynamics may contribute to understanding of the system's function.

In this paper we present a study of the hormone receptor binding/unbinding dynamics by steered molecular dynamics in order to clarify the role of some of the amino acid contacts that were identified in previous reports (Renaud et al., 1995; Ostrowski et al., 1998), and to identify the binding/unbinding pathways of the all-*trans* RA to/from its receptor.

Three possible unbinding pathways, shown in Fig. 2, were chosen based on an inspection of the liganded hRAR-γ structure. Given the importance of helices H11 and H12 in the binding process, as mentioned above, and the relative positioning of helix H12 in the available crystal structures, path 1 and path 2 for the unbinding of the hormone were chosen in the proximity of these helices. Selecting unbinding path 1 between helices H11 and H12 also allowed us to test the necessity of a displacement of helix H12 for binding. Following the same reasoning, unbinding path 2 was selected beneath helices H11 and H12. The third pathway, path 3, was chosen after examining the molecular surface of the protein. In the crystal structure of liganded hRAR-γ residue Lys-236 has two alternative conformations, one toward the interior, where it makes a salt bridge to the carboxylate end of t-RA, and one where the side chain points toward the exterior of the protein. The structure of the protein with Lys-236 in the latter conformation exhibits a "window" in the molecular surface that allows the carboxylate end of the hormone to be seen from the outside. Path 3 was chosen to study the unbinding of the hormone through this "window."

Which residues are most involved in the process of binding/unbinding? How large are the conformational changes induced in the protein by the binding/unbinding process? Can the hormone use one pathway for binding and another pathway for unbinding? These questions will be addressed in the present study using steered molecular dynamics (SMD) simulations (Leech et al., 1996; Grubmüller et al., 1996; Izrailev et al., 1997, 1998; Balsera et al., 1997; Isralewitz et al., 1997; Lüdemann et al., 1997; Stepaniants et al., 1997; Marrink et al., 1998; Wriggers and Schulten, 1998; Lu et al., 1998; Hermans et al., 1998) for unbinding t-RA from hRAR-γ. The time scale of the natural binding and unbinding of the hormone, of the order of milliseconds or longer, is unreachable for conventional molecular dynamics simulations which are limited to nanosecond time scales. In many cases, energy barriers involved in the binding or unbinding of a ligand to a receptor are too high for the ligand to cross the barrier spontaneously on a nanosecond time scale. SMD provides a means of accelerating the unbinding processes through application of external forces that lower the energy barriers and drive the ligand along its unbinding path on nanosecond time scales. By monitoring the forces applied and the response of the ligand, one can characterize the pathway and its intermediate states.

In the next section the methods used to simulate the unbinding of t-RA from hRAR-γ via different pathways are described. The Results section presents the response of the protein and of the hormone to the external forces during induced unbinding, and the Conclusions section summarizes arguments for the binding and unbinding pathways emerging from our study.

## METHODS

### Simulations

The simulations carried out were based on the x-ray crystallographic structure of the LBD of RAR bound to all-*trans* RA at 2 Å resolution (Renaud et al., 1995). All simulations were performed using the molecular dynamics program NAMD (Nelson et al., 1996) and version 22 of the CHARMM force field (Brooks et al., 1983; MacKerell, Jr. et al., 1992, 1998). For the charge distribution of the all-*trans* retinoic acid, Mulliken charges obtained using GAUSSIAN-94 (Frisch et al., 1995) at the Hartree-Fock level with a 6-31G* basis set, using the coordinates of heavy atoms from the crystal structure [entry 2lbd in the Protein Data Bank (Bernstein et al., 1977)] with hydrogens generated by the program QUANTA (MSI, 1994), were used. The equilibrium bond length, angles, torsional angles, and force constants for t-RA were derived from t-RA coordinates, following the treatment of retinal in bacteriorhodopsin as outlined in Humphrey et al. (1994) and force field parameters of molecules with similar chemical structure available in the CHARMM22 force field.

In all simulations we assumed a dielectric constant $\epsilon = 1$ and a cutoff of Coulomb forces with a switching function starting at 12 Å and reaching zero at a distance of 14 Å. All atoms, including hydrogens, were described explicitly. The hydrogen atom coordinates of both RAR and t-RA were generated using the HBUILD routine of X-PLOR (Brünger, 1992). An integration time step of 1 fs was employed.

The crystal structure of the LBD of RAR bound to all-*trans* RA (Renaud et al., 1995) contains 238 protein residues, the t-RA hormone, and 119 water molecules. To eliminate bad contacts and constraints due to crystal packing, the complex was energy-minimized for 500 steps of Powell algorithm in the presence of strong harmonic constraints on the $\alpha$-carbon atoms, followed by an additional 1500 steps without constraints. The protein-ligand complex was then immersed in the center of a 45-Å sphere of water molecules. Water for the solvation of the ligand-receptor complex was prepared as described in previous studies (Bishop et al., 1997; Kosztin et al., 1997). All water molecules that had one of the atoms closer than 1.8 Å to the protein or crystal water atoms were removed. Finally, only a 15-Å layer of water molecules around the protein was retained and all the other water molecules were deleted, resulting in a system of 14,574 atoms. The solvated protein-ligand system was equilibrated for 50 ps with velocity rescaling every 2.5 ps, followed by 60 ps of equilibration without rescaling. After equilibration, five separate simulation runs were conducted: two "free dynamics" simulations of 600 ps each with residue Lys-236 in both alternative positions and three SMD simulations. All simulations were carried out on 64 processors of a Cray T3E using 15 s of wall clock time per picosecond of simulation. The coordinates were saved every half a picosecond.

Preliminary simulation runs indicated that the part of the protein that does not contain the binding pocket, does not change its conformation during the unbinding of the hormone. Accordingly, during both stages of the equilibration and further simulations, soft harmonic constraints with a force constant of 0.5 kcal/mol Å$^2$ were applied to all the $\alpha$-carbon atoms in the top part of the protein (see Fig. 1 *c*). These constraints also served to prevent lateral movement of the protein during the simulated unbinding.

To induce unbinding of t-RA from the protein, the SMD approach was employed, i.e., an external force was applied to the hormone in a direction chosen along the selected unbinding path. Three different unbinding pathways, shown in Fig. 2, were explored, as outlined in the Introduction. The force was applied to the C3 atom of the $\beta$-ionone ring (see Fig. 8) in the simulations of hormone unbinding along path 1 and path 2, and to the C15 atom of the isoprene tail in simulations of unbinding along path 3. These atoms were harmonically restrained to a point moving with a constant velocity $v$ in a chosen direction. The hormone was thus pulled out of the protein. The external force exerted on the ligand was $\vec{F} = k(\vec{v}t - \vec{x})$, where $\vec{x}$ is the displacement of the restrained atom with respect to its original position, and $t$ is time elapsed from the beginning of the simulation. We chose a value of $k = 4$ kcal/mol Å$^2$ that is sufficiently small to render fluctuations of the applied force (due to thermal motion of the position $\vec{x}$ of the restrained atom) small relative to the magnitude of $\vec{F}$. At the same time, $k$ had to be sufficiently large so that the contraction of the harmonic "spring" due to the advancement of the ligand along the unbinding path by more than 1–2 Å resulted in a noticeable drop in the force (Izrailev et al., 1997; Balsera et al., 1997).

The strength of the adhesion of the ligand to the binding pocket depends on the velocity of pulling (Evans and Ritchie, 1997; Izrailev et al., 1997). The external force applied to the ligand lowers the energy barriers that the ligand has to surmount in order to unbind. Smaller velocities generally result in longer times and smaller forces required to induce unbinding because of the higher probability for the ligand to overcome the lowered energy barrier due to thermal fluctuations. Longer simulation times also lead to better sampling of the possible conformations and, therefore, to finding less resistive unbinding pathways. For each pathway suggested above a series of simulations with different pulling velocities $v$ and slightly different directions of the external force were carried out. In the analysis, however, only the simulations with the lowest velocity $v = 0.032$ Å/ps and the direction of the external force that resulted in the smallest distortion of the protein were included.

### Analysis

The atom selection commands and energy calculation routines available in X-PLOR were used to calculate nonbonded protein-ligand interaction energies as well as the magnitude of the external force exerted on the ligand. The structural deviations of the protein from the initial x-ray crystal structure were assessed on the basis of root mean square deviations (RMSD) and the analysis of the secondary structure. In computing the RMSD, the overall translational and rotational motions have been removed by superimposing the backbone of the protein in each configuration in all trajectories onto the backbone of the protein in the crystal structure using a least-square fitting algorithm (Kabsch, 1976).

The Debye-Waller factors, or B-factors, provide another important basis for comparing molecular dynamics trajectories with experimental results of x-ray crystallography. The theoretical temperature factors for the free dynamics of the ligand-protein system were computed according to

$$B = \frac{8\pi^2}{3} \langle (\Delta r)^2 \rangle, \tag{1}$$

where $\langle (\Delta r)^2 \rangle$ is the mean square of the atomic displacement averaged over the trajectories after a rigid body alignment against the coordinates of the initial structure. The average was taken over all non-hydrogen atoms in a given amino acid.

Direct, salt bridges, and water-mediated hydrogen bond interactions between the ligand and the protein residues were analyzed using the following conventions: two atoms were considered to form a hydrogen bond (A...H-D) if the acceptor-donor distance was <3.5 Å and if the A-H-D angle was between 120° and 180°; a salt bridge was considered to be formed by two residues with oppositely charged side chains within hydrogen bonding distance of each other.

## RESULTS

This section demonstrates the stability of the simulated protein-ligand systems, describes the response of the protein to the extraction of the hormone, and presents the measured adhesion force profiles as well as the interaction energies between the hormone and distinct residues along the unbinding pathways.

## Free dynamics

The ligand bound system of hRAR-γ was simulated for 600 ps with residue Lys-236 in both alternative conformations, as described in Methods. The RMSD of the protein atoms from their crystal structure positions are presented in Fig. 3. The results suggest that the overall structure of the system was well preserved in both simulations. RMSD values were normally below 3 Å, only the C-terminal end of H11 exhibited RMSD values between 3 and 3.5 Å (data not shown).

Fig. 4 compares observed (Renaud et al., 1995) and simulated crystallographic B-factors. The atomic fluctuations in both simulations match the experimental B factor pattern relatively well. Differences occurred in the segments 205:240, 295:313, 390:420, where the calculated fluctuations significantly exceed the experimental ones. The 205:240 stretch corresponds to a coil region and the N-terminal end of helix H3. The 295:313 region corresponds to one of the β-sheets and helix H6 and the 390:420 region corresponds to the C-terminal end of helix H11 and the coil that connects it to helix H12. All three regions lie at the surface of the protein and exhibit strong interactions with the solvent, which may explain a higher mobility than in the crystal.

The binding pocket did not deform during the free dynamics simulations and the ligand maintained its position and orientation, fluctuating around its equilibrium position. This is expected because the packing around t-RA is relatively tight. Analysis of the hydrogen bonding network between the hormone and the protein residues confirmed that the hydrogen bonding network observed in the crystal structure was well maintained during the simulations.

The t-RA ligand is completely buried in the protein interior. There are no openings in the molecular surface of the protein to connect the active site to the protein surface, except when residue Lys-236 is oriented toward the outside of the protein. As shown in Fig. 5, after 600 ps of dynamics this opening became larger and allowed a better view of the ligand inside the binding pocket. Large fluctuations of res-idue Lys-236 in both simulations allowed water molecules to enter the binding pocket and cluster around the carboxylate end of the hormone. Some of these water molecules moved between residues Arg-278, Arg-274, Ser-289, Lys-236, and the carboxylate end of the hormone. This suggests that water may destabilize salt bridges between the hormone and the residues lining the opening, leading to a widening of the opening in the molecular surface. Interestingly, in the simulation with Lys-236 oriented toward the binding pocket and making a contact to the carboxylate group of the hormone, no opening was discernible at the beginning of the simulation; nevertheless, a "window" in the molecular surface of the protein developed as the simulation proceeded. Even though in this case residue Lys-236 was hydrogen-bonded to the carboxylate end of t-RA, its fluctuations around the equilibrium position were still large.

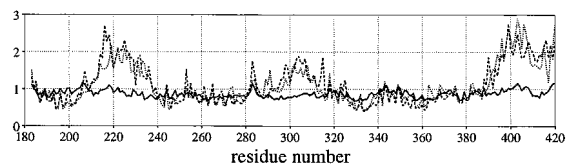

FIGURE 4 Fluctuations of protein side chains computed, as described in Methods, from free dynamics simulations (*dashed line* corresponding to Lys-236 oriented inward and *dotted line* corresponding to Lys-236 oriented outward) and from crystallographic data (Renaud et al., 1995) (*solid line*).

## Unbinding along path 1

The force required to extract the hormone from the binding pocket (see Methods) along path 1 is shown in Fig. 6. The simulation revealed distinct features of the unbinding process. Throughout the course of the unbinding, residues altered their interactions with the ligand. For example, analysis of the hydrogen bonds showed a sequence of contacts between ligand and side groups lining the pathway that are made and broken during unbinding. The interaction energies between the ligand and the protein were calculated separately for the β-ionone ring and for the isoprene tail. The isoprene tail contribution is mainly electrostatic, whereas the largest contribution to the total interaction energy for the β-ionone ring arose from many small vdW interactions.

During the first 170 ps of unbinding along path 1 a steady increase of the applied force was observed, as shown in Fig. 6. After 170 ps, the hydrogen bond made by the tail of the hormone to one of the amino nitrogens of Arg-278 was broken, while the bond made to the other amino nitrogen of Arg-278 was weakened. This event was manifested by a sharp increase of the interaction energy of the hormone with Arg-278, as shown in Fig. 7, and by a decrease of the applied force by ~70 pN.

Between 170 and 270 ps of path 1 unbinding, the hormone had to overcome vdW interactions with residues obstructing the exit and electrostatic interactions of the isoprene tail with residues anchoring it in the binding pocket. Residues Gly-393, Ala-397, Leu-400, Met-408, and Leu-416 were in close vdW contact with the β-ionone ring of the hormone at the beginning of the simulation. The hormone had to push past these residues to follow along path 1. At 200 ps, the β-ionone ring of the hormone rotated ~35° around the C6-C7 bond (see Fig. 8) with respect to the isoprene tail. X-ray crystallographic, NMR, and theoretical studies of retinoids have shown that rotation around the C6-C7 bond is one of the most interesting properties of retinoids (van Aalten et al., 1996). We have monitored, therefore, the evolution of the C1-C6-C7-C8 dihedral angle, presented in Fig. 8. Rotation of the β-ionone ring brought residues Gly-393, Leu-416, and Ile-412 in close contact with the 5-methyl group of the ring and residues Ala-397, Leu-400, and Met-408 in close contact with the geminal methyl groups of the hormone. The force needed for extraction remained nearly constant during this time, after which there was a sharp increase in its magnitude by ~100 pN.

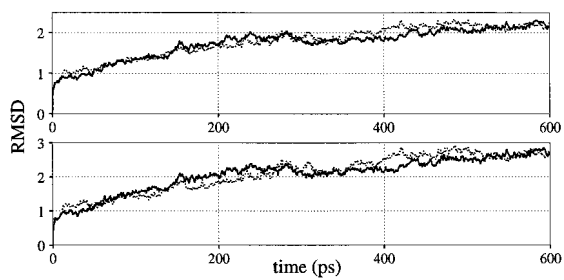

FIGURE 3 *Top*: RMS deviation for all heavy atoms of (hRAR)-γ in both free dynamics simulations; *bottom*: RMS deviation for all heavy atoms of (hRAR)-γ in the bottom part of the protein where no α-carbon atoms were restrained. In both graphs the continuous line represents the system with residue Lys-236 making contact with the t-RA hormone (i.e., oriented inward), and the dotted line represents the system with Lys-236 residue oriented toward the solvent.

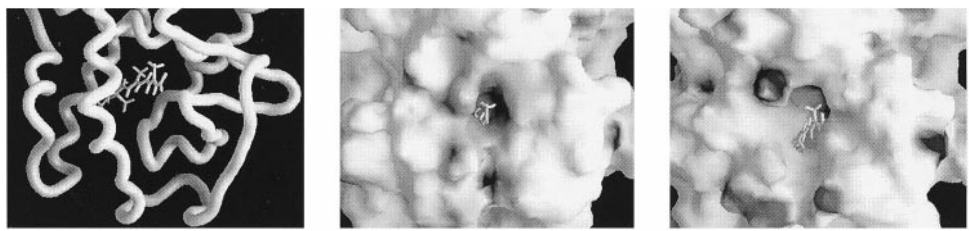

FIGURE 5 *Left*: backbone representation of the bottom half of the protein showing the orientation of the hormone inside the binding pocket; *center*: opening in the molecular surface of hRAR-γ with Lys-236 oriented toward the solvent, i.e., outward; *Right*: the "window" present at the beginning of the simulation becomes larger after 600 ps and allows a better view of the hormone inside the binding pocket. Figure created using GRASP (Nicholls et al., 1991).

At 270 ps, the carboxylate end of the hormone was still hydrogen-bonded to residues Lys-236, Ser-289, and Phe-288. A sharp increase in the applied force to ~540–580 pN was necessary to break these bonds. The breaking of these hydrogen bonds takes place between 270 and 320 ps of path 1 unbinding. At the same time, the 5-methyl group of the β-ionone ring repositioned itself between the side chains of Ile-412 and Leu-416, pushing both aside. The geminal methyl groups were still behind residues Leu-400 and Met-408. These residues were finally pushed aside at ~360 ps. At this point, there were no more residues obstructing the movement of the β-ionone ring of the hormone along path 1; consequently, the force decreased by ~150 pN. The following increase in the force, between 400 and 430 ps, was due to the 19-methyl group on the isoprene tail (see Fig. 8) coming into close contact with Leu-400. At 445 ps the 19-methyl group slipped past this residue and the β-ionone ring found itself almost completely outside the binding pocket.

After ~450 ps the β-ionone ring had passed between helices H11 and H12 and moved into the solvent. Passing of the isoprene tail between the two helices required a force of ~170 pN. After 720 ps of simulation an increase in the pulling force was observed for ~100 ps. Analysis of the interaction energy between protein residues along the unbinding pathway and the hormone revealed that the increase in the applied force was due to the interaction between the carboxylate end of the hormone and Arg-396 first, and then Arg-413 (see Fig. 7). After 1 ns of simulation, the hormone, completely out of the binding pocket, still maintains the hydrogen bond to Arg-413.

The simulations showed that perturbation of the salt bridge between residues Glu-414 and Lys-264 was not necessary for the unbinding of the hormone from the active site. Relatively small and localized displacements and rotations of protein side chains, involving an RMSD of 3–4 Å for the C-terminal end of helix H11 and the coil that connects it to helix H12, were sufficient to allow the t-RA hormone to leave the binding pocket. These deformations and displacements mainly affected the C-terminal end of helix H11 and the coil that connects it to helix H12, whereas the secondary structure of helix H12 was very well maintained. We note that during the free dynamics simulations this region of the protein displayed high fluctuations, which may account for some of the changes in the structure during the forced unbinding. Movement of the hormone along path 1 led to inclusion of water molecules in the binding pocket through the region surrounding the carboxylate end of the hormone. One water molecule, hydrogen-bonded to the carboxylate end of the hormone, follows the hormone along the entire unbinding path 1.

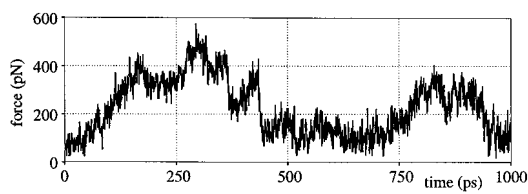

FIGURE 6 Force required to extract the hormone along path 1.

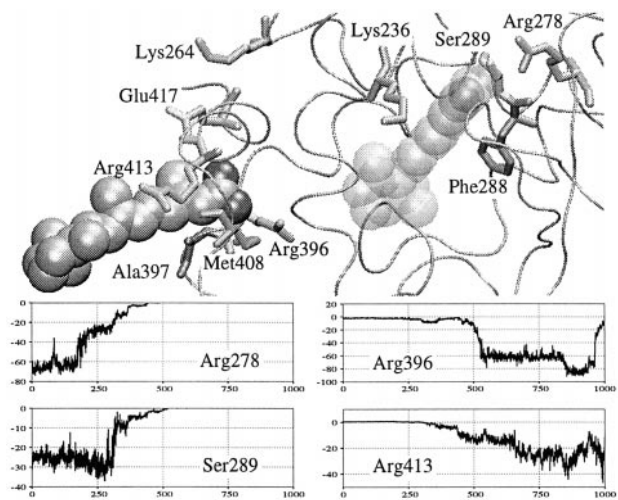

FIGURE 7 *Top*: Snapshot, at 750 ps, showing the hormone leaving the binding pocket along path 1 and the location of key amino acids; the initial position of the hormone is represented by transparent vdW spheres; *bottom*: time evolution of the interaction energies (in kcal/mol) between some of the residues along the unbinding pathway and the t-RA hormone. The electrostatic interaction of the carboxylate end of the hormone with residues Arg-278 and Ser-289 decreases as the interaction with residues Arg-396 and Arg-413 increases.

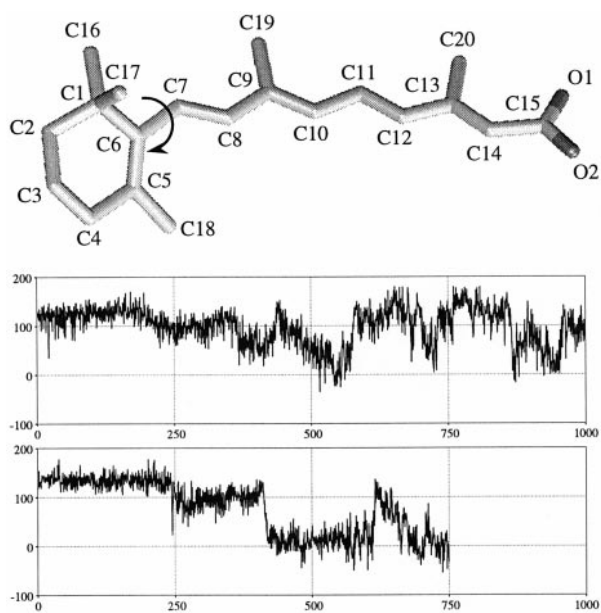

FIGURE 8 *Top*: crystal structure of the all-*trans* retinoic acid hormone. The arrow indicates torsion around the C6-C7 bond rotating the β-ionone ring. *Bottom*: time evolution of the C1-C6-C7-C8 dihedral angle of the t-RA hormone during the forced unbinding along path 1 (*top*) and path 3 (*bottom*); change of the angle implies rotation of the hormone β-ionone ring.

### Unbinding along path 2

Forced unbinding along path 2, depicted in Fig. 2, proved to strongly affect protein conformation. Various directions for the applied force were tried to avoid conformational changes, but without success. Several residues had to be dislodged in order for the hormone to leave the binding pocket; residues Trp-227 and Phe-230 had to be completely reoriented. The C-terminal end of helix 3 unraveled, some of the residues in this region trailing the hormone.

Despite the considerable rearrangement of protein side groups, the force required to move the ligand along path 2, shown in Fig. 9, did not exceed 460 pN. However, the force decreased only after ~470 ps of simulated unbinding. Before this occurred the hormone experienced vdW interactions with many of the residues along the unbinding pathway. These vdW interactions were so strong that the breaking of the hydrogen bonding network between protein residues and the hormone carboxylate end was not an event that could be attributed to an increase or decrease in the force. After ~500 ps many of the residues along the pathway had reoriented, and the β-ionone ring emerged from the interior of the protein. After 750 ps of simulation time, the t-RA hormone was still partially inside the protein and marked deformations of the protein were still present.

### Unbinding along path 3

The t-RA binding site "window" mentioned above, as well as the high temperature factors of the same region in the TR LBD structure (Wagner et al., 1995), prompted us to extract t-RA along path 3. Charged residues surrounding the carboxylate end of the ligand form a ring around the binding site "window" (see Fig. 10) and act as an anchor for the hormone. The force required to pull the hormone along path 3 is presented in Fig. 11.

Initially, the unbinding of the hormone proceeded slowly, since the hormone remained tightly bound while the applied force steadily increased. Forces of over 500 pN were required to surmount the high electrostatic barrier imposed by residues around the carboxylate end of the hormone. The first peak in the force, of 520 pN, at ~170 ps, corresponded to the breaking of the salt bridge between the carboxylate end of the hormone and Arg-274. The force decreased by ~180 pN during the following 20 ps before starting to increase again at 190 ps, in order to break the hydrogen bonds and salt bridges between the retinoic acid and residues Lys-229, Ser-289, and Phe-288. Consequently, another peak of 530 pN occurred at 240 ps. Breaking of the salt bridge between the hormone and Arg-278 at 260 ps required a force of ~470 pN. The abrupt decrease of the interaction energy between the hormone and residues Arg-274, Arg-278, Lys-229, Lys-236, Ser-289, and Phe-288 correlated with the changes of the applied force (see Fig. 11).

The β-ionone ring of the hormone was initially in close contact with residues Met-272, Phe-304, Leu-268, Leu-271, and Met-415. After 245 ps of path 3 unbinding, the move-ment of the hormone as well as the interaction with the surrounding residues led to a slight rotation of the $\beta$-ionone ring around the C6-C7 bond, as shown in Fig. 8. In the new orientation, the bulky methyl groups of the $\beta$-ionone ring were not occluded by any residue side chains for $\sim$60 ps. At 320 ps the geminal methyl groups came in close contact with residues Ala-234 and Leu-233, leading to an increase in force of $\sim$100 pN. At 410 ps, another rotation of the $\beta$-ionone ring around the C6-C7 bond accompanied the surmounting of this obstacle and the force dropped by $\sim$200 pN. At this point the $\beta$-ionone ring was almost out of the binding pocket with no residues obstructing the further pathway. Residues still in close contact with the hormone led to a slight increase of the force, between 425 and 475 ps. At 475 ps the drop in force signaled the completed unbinding of the hormone. From this point on, the entire hormone moved in solvent; the simulations were stopped at 750 ps.

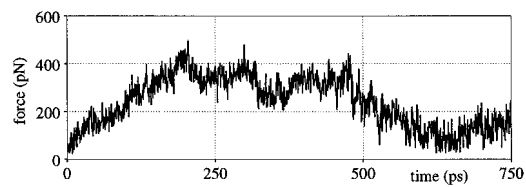

FIGURE 9 Force required to extract the hormone along path 2.

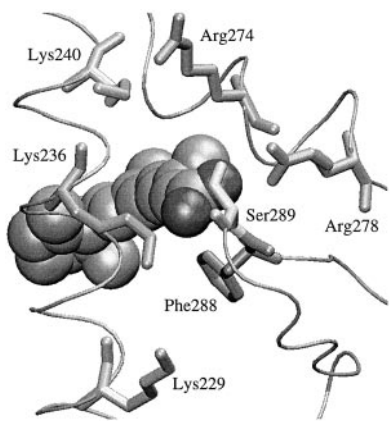

FIGURE 10 Protein residues located within 5 Å of the carboxylate end of t-RA are lining the "window" in the molecular surface.

The secondary structure of the protein was well maintained during the unbinding along path 3. The applied force affected only helix H3 around the Lys-236 residue, inducing oscillations of residue Lys-236. However, this region also exhibited strong fluctuations in the simulations of the protein-ligand system without external forces applied. Water molecules, clustered around the carboxylate end of the hormone, entered the binding pocket, and remained clustered around the "window" mentioned above.

## CONCLUSIONS

Three different pathways for the unbinding process of t-RA from RAR were explored. Simulation results indicated that it is possible to unbind the hormone along path 1 and path 3 without greatly affecting the structure of the protein. Particular characteristics of the simulated pathways discussed below suggest path 1 as the *binding pathway* for the hormone and path 3 as the *unbinding pathway*.

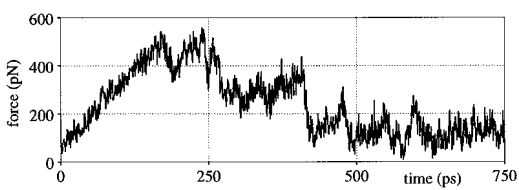

FIGURE 11 Force required to extract the hormone along path 3.

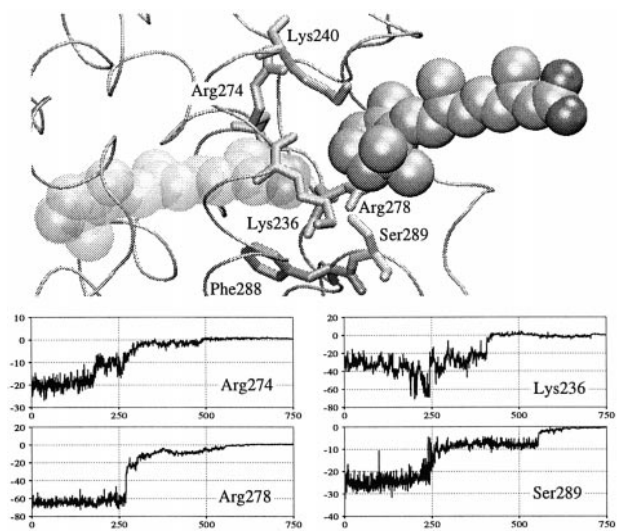

FIGURE 12 *Top*: snapshot, at 600 ps, showing the hormone leaving the binding pocket along path 3 and the location of key amino acids; the initial position of the hormone is represented by transparent vdW spheres; *bottom*: time evolution of the energies (in kcal/mol) of interaction between the t-RA hormone and charged residues that line the "window."

The force profile for the unbinding of t-RA from its receptor showed similar characteristics along paths 1 and 3. The measured force gradually increased toward its maximum value after which it decreased as the hormone was leaving the binding pocket. Even though in these two simulations the hormone was pulled out of the binding pocket using a force applied to opposite ends of the hormone, two important steps were observed in both simulations. One step involved breaking of the salt bridges between the carboxylate end of the hormone and charged or polar residues that surround it, the other step involved overcoming the van der Waals interactions between protein residues and the $\beta$-ionone ring of the hormone. This two-step scenario can explain why path 1 may be considered the *binding pathway* and path 3 the *unbinding pathway* of the hormone.

Let us briefly summarize, for this purpose, the characteristics of path 1 unbinding. The $\beta$-ionone ring of the hormone had to pass through the residues that close the entrance to the binding pocket, and the hydrogen bonding network between the carboxylate end and protein residues had to be broken. In case of path 1 these two events develop almost simultaneously. When the hormone leaves the binding pocket, the electrostatic interaction between the hormone and residues inside the binding pocket is replaced by the interactions with charged residues on the surface of the protein.

Now let us suppose that path 1 is the *binding pathway* for the hormone and that helix H12 for the apo form of hRAR-$\gamma$

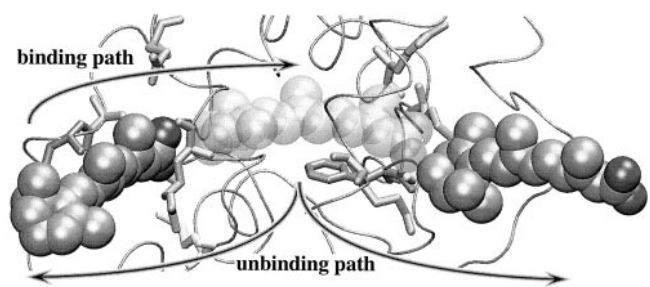

FIGURE 13　Representation of the superimposed binding/unbinding events (bound position of the hormone shown as transparent vdW spheres). Residues lining the binding/unbinding windows, as shown in Figs. 7 and 12, are represented in licorice.

is folded in the same position as in the crystal structure of the holo hRAR-$\gamma$ (Renaud et al., 1995). Then, using the reversed order of events observed in the simulated unbinding along path 1, one can describe the binding mechanism of the hormone. First Arg-413, and then Arg-396, attract and orient the carboxylate end of the hormone toward the binding pocket. When the hormone is within hydrogen bonding distance of these two residues, it also experiences the influence of the charged and polar residues Arg-278, Lys-236, Arg-274, and Ser-289 located at the opposite end of the binding pocket. The two steps described above take place almost simultaneously, i.e., the strong electrostatic attraction between the carboxylate end of the hormone and residues Arg-278, Lys-236, Ser-289, and Arg-274 helps the $\beta$-ionone ring to pass between protein residues, thus leading to the penetration of the hormone into the binding pocket. The entrance of path 1 is surrounded by highly fluctuating residues. These residues may become more ordered upon binding of the hormone, making contacts with the hormone or with other protein residues. Induced ordering of the protein side chains may be favorable for the hormone entry without requiring a motion of helix H12.

Study of the unbinding of the hormone along path 3 was prompted by the existence of the "window" in the molecular surface that allows access to the binding site, as shown in Fig. 5. Interestingly, this "window" is lined with charged and polar residues that clearly have an important role in attracting and anchoring the hormone into the binding pocket. There are approximately no changes in the structure of the protein upon unbinding, the applied force is not larger along path 3 than along path 1, and the two-step process is discernible for path 3 as well.

There are two arguments that make path 3 an unlikely candidate for a binding pathway. First, the carboxylate end of the hormone should be the one to be attracted by the point of entry into the protein since it furnishes stronger and more specific interactions than the $\beta$-ionone ring. However, if path 3 represented a binding path, the $\beta$-ionone ring had to go through the "window" into the binding pocket first. The "window" would need to be sufficiently large to let the bulky ring pass through. However, the "window" explored in our simulations was too small to allow ligand access, except after reordering of residue Lys-236 and after hydration events.

One may then consider path 3 as a possible *unbinding pathway*, either as the sole unbinding pathway or an alternative to path 1. There is no experimental evidence that the hormone leaves the binding pocket the same way it entered. In fact, conformational changes of the RAR ligand binding domain could easily render the binding of the hormone along path 1 irreversible on relevant time scales. The dissociation of the hormone along path 3 would be a slow process, and thermal fluctuations that govern the unbinding may cause a sufficient opening of the "window" for the hormone to pass through. Presence of ions around that region in the protein as well as intrusion of water molecules in the binding pocket may help destabilize the hydrogen bonding network between the carboxylate end of the hormone and protein residues and, thus, lead to the opening of the binding pocket.

One might criticize the simulations presented in this paper for the choices made in modeling the ligand binding/unbinding process: preselected direction for the applied force, short simulation time, force field choice, and force constant choice for the harmonic spring. However, it was shown previously (Grubmüller et al., 1996; Izrailev et al., 1997; Isralewitz et al., 1997; Lüdemann et al., 1997; Stepaniants et al., 1997; Marrink et al., 1998; Lu et al., 1998; Hermans et al., 1998) that despite some shortcomings, results obtained from SMD simulations yield important qualitative insights into binding mechanisms and correlate well with experimental data.

Variations in the choice of the force constant for the harmonic spring as well as different pulling rates influence the results, as outlined in Methods. Through model calculations it was shown that by using soft springs, one can better measure the global properties of the system than when using stiff springs (Izrailev et al., 1997; Balsera et al., 1997). Simulations with different pulling velocities $v$ showed that forces required for the unbinding of hormone are higher for larger values of $v$ than for smaller ones. As $v$ decreases, the ligand has more time to sample the conformational space and to search for a path of least resistance along the chosen direction. In the simulations the sampling times are limited to nanoseconds, which is not sufficient to find the best possible path, and such a path must be preselected. Nevertheless, the hormone can still adjust its conformation and position in such a way that crossing of very large energy barriers is avoided.

It was observed in the simulations that the protein response does not change significantly with the changes in the velocity $v$. For example, the deformations induced in the C-terminal end of helix H11 by the forced unbinding of the hormone appear in simulations along path 1 with $v = 0.032$ Å/ps as well as with $v = 0.08$ Å/ps. These deformations may be due to the inability of the protein to adequately respond to the perturbation on the time scale of the simulations even when small values of $v$ are employed, as well as due to the intrinsic flexibility of the protein in that region.

Even though the magnitude of the measured forces is influenced by the choice of force constant for the harmonic spring or the pulling velocity, the qualitative features of the unbinding process remain unchanged. Based on this, we suggest that the results of our SMD simulations are qualitatively representative for the binding/unbinding process and applicable to other members of the nuclear hormone receptors family.

Retinoids, the natural and synthetic derivatives of vitamin A, are crucial for cell differentiation and proliferation and embryonic development. Each cell type manufactures its own pool of retinoids, and these interact with retinoic acid receptors in the cell nucleus to modulate gene transcription. Because of their ability to stimulate normal cell differentiation, retinoids have been used to treat cutaneous T-cell lymphomas, leukoplakia, squamous cell carcinomas of the skin, and basal cell carcinomas. One of the retinoids, all-*trans* retinoic acid, has recently received FDA approval for two separate uses: the topical treatment of photoaged skin and the oral treatment of acute promyelocytic leukemia. Treatment of diseases with retinoic acid causes unwanted side effects, such as fever, respiratory distress, and hypotension. In order to eliminate these side effects, other ligands with high affinity to the receptor should be designed. Knowledge of ligand binding pathways can provide additional information, such as whether a ligand is likely to be able to enter the binding pocket, that helps ligand screening and accelerate drug design. Also, as the unbinding of the hormone represents the path of least resistance, studying the early stages of the unbinding can reveal portions of the receptor that are usually flexible. This flexibility may be exploited in the design of novel ligands that open up new pockets which may not be present in the absence of the novel ligand.

The authors thank J. Katzenellenbogen for helpful discussions and suggestions, R. Brunner and J. Phillips for help with NAMD, and F. Molnar for a critical reading of the manuscript.

This work was supported by National Institutes of Health Grant PHS 5 P41 RR05969-04, National Science Foundation Grants BIR-9318159 and BIR 94-23827 EQ, a grant from the Roy J. Carver Charitable Trust, and by the MCA 93S028P computer time grant at Pittsburgh Supercomputing Center.

The figures in this paper were created with the molecular graphics program VMD (Humphrey et al., 1996) (http://www.ks.uiuc.edu/Research/VMD).

## REFERENCES

Andersson, M. L., K. Nordström, S. Demczuck, M. Harbers, and B. Vennström. 1992. Thyroid hormone alters the DNA binding properties of chicken thyroid hormone receptors $\alpha$ and $\beta$. *Nucleic Acids Res.* 20:4803–4810.

Balsera, M., S. Stepaniants, S. Izrailev, Y. Oono, and K. Schulten. 1997. Reconstructing potential energy functions from simulated force-induced unbinding processes. *Biophys. J.* 73:1281–1287.

Bernstein, F. C., T. F. Koetzle, G. J. Williams, E. F. Meyer, M. D. Brice, J. R. Rogers, O. Kennard, T. Shimanouchi, and M. Tasumi. 1977. The Protein Data Bank: a computer-based archival file for macromolecular structures. *J. Mol. Biol.* 112:535–542.

Bishop, T. C., D. Kosztin, and K. Schulten. 1997. How hormone receptor-DNA binding affects nucleosomal DNA: the role of symmetry. *Biophys. J.* 72:2056–2067.

Bourguet, W., M. Ruff, P. Chambon, H. Gronemeyer, and D. Moras. 1995. Crystal structure of the ligand-binding domain of the human nuclear receptor RXR-$\alpha$. *Nature.* 375:377–382.

Brent, G. A., M. K. Dunn, J. W. Harney, T. Gulick, and P. R. Larsen. 1989. Thyroid hormone aporeceptor represses $T_3$ inducible promoters and blocks activity of the retinoic acid receptor. *New Biol.* 1:329–336.

Brooks, B. R., R. E. Bruccoleri, B. D. Olafson, D. J. States, S. Swaminathan, and M. Karplus. 1983. CHARMM: a program for macromolecular energy, minimization, and dynamics calculations. *J. Comp. Chem.* 4:187–217.

Brünger, A. T. 1992. X-PLOR, Version 3.1: A System for X-Ray Crystallography and NMR. The Howard Hughes Medical Institute and Department of Molecular Biophysics and Biochemistry, Yale University.

Brzozowski, A. M., A. C. W. Pike, Z. Dauter, R. E. Hubbard, T. Bonn, O. Engström, L. Öhman, G. L. Greene, J. A. Gustafsson, and M. Carlquist. 1997. Molecular basis of agonism and antagonism in the oestrogen receptor. *Nature.* 389:753–758.

Damm, K., C. C. Thompson, and R. M. Evans. 1989. Protein encoded by v-*erb*A functions as a thyroid-hormone receptor antagonist. *Nature.* 339:593–597.

Driscoll, J. E., C. L. Seachord, J. A. Lupisella, R. P. Darveau, and P. R. Reczek. 1996. Ligand-induced conformational changes in the human retinoic acid receptor $\gamma$ detected using monoclonal antibodies. *J. Biol. Chem.* 271:22969–22975.

Evans, E., and K. Ritchie. 1997. Dynamic strength of molecular adhesion bonds. *Biophys. J.* 72:1541–1555.

Freedman, L. P., and B. F. Luisi. 1993. On the mechanism of DNA binding by nuclear hormone receptors: a structural and functional perspective. *J. Cell. Biochem.* 51:140–150.

Frisch, M. J., G. W. Trucks, H. B. Schlegel, P. M. W. Gill, B. G. Johnson, M. A. Robb, J. R. Cheeseman, T. Keith, G. A. Petersson, J. A. Montgomery, K. Raghavachari, M. A. Al-Laham, V. G. Zakrzewski, J. V. Ortiz, J. B. Foresman, C. Y. Peng, P. Y. Ayala, W. Chen, M. W. Wong, J. L. Andres, E. S. Replogle, R. Gomperts, R. L. Martin, D. J. Fox, J. S. Binkley, D. J. Defrees, J. Baker, J. P. Stewart, M. Head-Gordon, C. Gonzalez, and J. A. Pople. 1995. Gaussian 94, Revision b.3. Gaussian Inc., Pittsburgh, PA.

Grubmüller, H., B. Heymann, and P. Tavan. 1996. Ligand binding and molecular mechanics calculation of the streptavidin-biotin rupture force. *Science.* 271:997–999.

Hermans, J., G. Mann, L. Wang, and L. Zhang. 1998. Simulation studies of protein-ligand interactions. *In* Algorithms for Macromolecular Modelling. P. Deuflhard, J. Hermans, B. Leimkuhler, A. Mark, R. D. Skeel, and S. Reich, editors. Lecture Notes in Computational Science and Engineering. Springer-Verlag, New York. In press.

Humphrey, W. F., A. Dalke, and K. Schulten. 1996. VMD—visual molecular dynamics. *J. Mol. Graphics.* 14:33–38.

Humphrey, W., I. Logunov, K. Schulten, and M. Sheves. 1994. Molecular dynamics study of bacteriorhodopsin and artificial pigments. *Biochemistry.* 33:3668–3678.

Isralewitz, B., S. Izrailev, and K. Schulten. 1997. Binding pathway of retinal to bacterio-opsin: a prediction by molecular dynamics simulations. *Biophys. J.* 73:2972–2979.

Izrailev, S., S. Stepaniants, M. Balsera, Y. Oono, and K. Schulten. 1997. Molecular dynamics study of unbinding of the avidin-biotin complex. *Biophys. J.* 72:1568–1581.

Izrailev, S., S. Stepaniants, B. Isralewitz, D. Kosztin, H. Lu, F. Molnar, W. Wriggers, and K. Schulten. 1998. Steered molecular dynamics. *In* Algorithms for Macromolecular Modelling. P. Deuflhard, J. Hermans, B. Leimkuhler, A. Mark, R. D. Skeel, and S. Reich, editors. Lecture Notes in Computational Science and Engineering. Springer-Verlag, New York. In press.

Judelson, C., and M. L. Privalsky. 1996. DNA recognition by normal and oncogenic thyroid hormone receptors—unexpected diversity in half-site specificity controlled by non-zinc-finger determinants. *J. Biol. Chem.* 271:10800–10805.

Kabsch, W. 1976. A solution for the best rotation to relate two sets of vectors. *Acta Crystallogr. A.* 32:922–923.

Klaholz, B. P., J. P. Renaud, A. Mitschler, C. Zusi, P. Chambon, H. Gronemeyer, and D. Moras. 1998. Conformational adaptation of agonist to the human nuclear receptor rarγ. *Nature Struct. Biol.* 5:199–202.

Kosztin, D., T. C. Bishop, and K. Schulten. 1997. Binding of the estrogen receptor to DNA: the role of waters. *Biophys. J.* 73:557–570.

Krust, A., S. Green, P. Argos, V. Kumar, P. Walter, J. Bornert, and P. Chambon. 1986. The chicken oestrogen receptor sequence: homology with v-*erb*A and the human oestrogen and glucocorticoid receptors. *EMBO J.* 5:891–897.

Kumar, V., S. Green, G. Stack, M. Berry, J. R. Jin, and P. Chambon. 1987. Functional domains of the human estrogen receptor. *Cell.* 51:941–951.

Lee, J. W., F. Ryan, J. C. Swaffield, S. A. Johnston, and D. D. Moore. 1995. Interaction of thyroid-hormone receptor with a conserved transcriptional mediator. *Nature.* 374:91–94.

Leech, J., J. Prins, and J. Hermans. 1996. SMD: visual steering of molecular dynamics for protein design. *IEEE Comp. Sci. & Eng.* 3:38–45.

Lu, H., B. Isralewitz, A. Krammer, V. Vogel, and K. Schulten. 1998. Unfolding of titin immunoglobulin domains by steered molecular dynamics simulation. *Biophys. J.* 75:662–671.

Lüdemann, S. K., O. Carugo, and R. C. Wade. 1997. Substrate access to cytochrome P450cam: a comparison of a thermal motion pathway analysis with molecular dynamics simulation data. *J. Mol. Model.* 3:369–374.

MacKerell, Jr., A. D., D. Bashford, M. Bellott, R. L. Dunbrack, Jr., J. Evanseck, M. J. Field, S. Fischer, J. Gao, H. Guo, S. Ha, D. Joseph, L. Kuchnir, K. Kuczera, F. T. K. Lau, C. Mattos, S. Michnick, T. Ngo, D. T. Nguyen, B. Prodhom, I. W. E. Reiher, B. Roux, M. Schlenkrich, J. Smith, R. Stote, J. Straub, M. Watanabe, J. Wiorkiewicz-Kuczera, D. Yin, and M. Karplus. 1998. All-hydrogen empirical potential for molecular modeling and dynamics studies of proteins using the CHARMM22 force field. *J. Phys. Chem. B.* 102:3586–3616.

MacKerell, Jr., A. D., D. Bashford, M. Bellott, R. L. Dunbrack, Jr., J. Evanseck, M. J. Field, S. Fischer, J. Gao, H. Guo, S. Ha, D. Joseph, L. Kuchnir, K. Kuczera, F. T. K. Lau, C. Mattos, S. Michnick, T. Ngo, D. T. Nguyen, B. Prodhom, B. Roux, M. Schlenkrich, J. Smith, R. Stote, J. Straub, M. Watanabe, J. Wiorkiewicz-Kuczera, D. Yin, and M. Karplus. 1992. Self-consistent parametrization of biomolecules for molecular modeling and condensed phase simulations. *FASEB J.* 6:143a. (Abstr.).

Marrink, S.-J., O. Berger, P. Tieleman, and F. Jähnig. 1998. Adhesion forces of lipids in a phospholipid membrane studied by molecular dynamics simulations. *Biophys. J.* 74:931–943.

Martinez, E., D. D. Moore, E. Keller, D. Pearce, V. Robinson, P. Macdonald, S. S. Simons, E. Sanchez, and M. Danielsen. 1997. The nuclear receptor resource project. *Nucleic Acids Res.* 25:163–165.

MSI. 1994. Quanta 4.0. Molecular Simulations Inc., Burlington, Massachusetts.

Nelson, M., W. Humphrey, A. Gursoy, A. Dalke, L. Kalé, R. D. Skeel, and K. Schulten. 1996. NAMD—a parallel, object-oriented molecular dynamics program. *J. Supercomputing App.* 10:251–268.

Nicholls, A., K. A. Sharp, and B. Honig. 1991. Protein folding and association: insights from the interfacial and thermodynamic properties of hydrocarbons. *Proteins: Struct., Funct., Genet.* 11:281–296.

Ostrowski, J., T. Roalsvig, L. Hammer, A. Marinier, J. E. S. Jr., K. L. Yu, and P. R. Reczek. 1998. Serine 232 and methionine 272 define the ligand binding pocket in retinoic acid receptor subtypes. *J. Biol. Chem.* 273:3490–3495.

Renaud, J. P., N. Rochel, M. Ruff, V. Vivat, P. Chambon, H. Gronemeyer, and D. Moras. 1995. Crystal structure of the RAR-γ ligand-binding domain bound to all-*trans* retinoic acid. *Nature.* 378:681–689.

Stepaniants, S., S. Izrailev, and K. Schulten. 1997. Extraction of lipids from phospholipid membranes by steered molecular dynamics. *J. Mol. Model.* 3:473–475.

Tanenbaum, D. M., Y. Wang, S. P. Williams, and P. B. Sigler. 1998. Crystallographic comparison of the estrogen and progesterone receptor's ligand binding domains. *Proc. Natl. Acad. Sci. USA.* 95:5998–6003.

van Aalten, D. M. F., B. L. de Groot, H. Berendsen, and J. B. C. Findlay. 1996. Conformational analysis of retinoids and restriction of their dynamics by retinoid-binding proteins. *Biochem. J.* 319:543–550.

Vombaur, E., M. Harbers, S. J. Um, A. Benecke, P. Chambon, and R. Losson. 1998. The yeast ADA complex mediates the ligand-dependent activation function AF-2 of retinoid X and estrogen receptors. *Genes Dev.* 12:1278–1289.

Wagner, R., J. W. Apriletti, M. E. McGrath, B. L. West, J. D. Baxter, and R. J. Fletterick. 1995. A structural role for hormone in the thyroid hormone receptor. *Nature.* 378:690–697.

Williams, S. P., and P. B. Sigler. 1998. Atomic structure of progesterone complexed with its receptor. *Nature.* 393:392–396.

Wriggers, W., and K. Schulten. 1998. Investigating a back door mechanism of actin phosphate release by steered molecular dynamics. *Proteins Struct. Funct. Genet.* (in press).